\documentclass[eqsecnum,aps,epsf]{revtex4}

\usepackage{graphicx}
\usepackage{bm}
\usepackage{theorem}
\usepackage{here}

\newcommand{\square}{\kern1pt\vbox{\hrule height  1.2pt\hbox{\vrule
width 1.2pt\hskip 3pt
\vbox{\vskip 6pt}\hskip  3pt\vrule width 0.6pt}\hrule height
0.6pt}\kern1pt}
\def\Vec#1{\mbox{\boldmath $#1$}}

\begin{document}

\title{Chaos of Yang-Mills Field in  Class  A Bianchi Spacetimes}

\author{Yoshida Jin$^{1}$ 
\footnote{{Electronic address:jin@gravity.phys.waseda.ac.jp}}}
\author{Kei-ichi Maeda$^{1,2,3}$ 
\footnote{Electronic address:maeda@gravity.phys.waseda.ac.jp}\\~}
\affiliation{$^{1}$Department of Physics, Waseda University, 3-4-1 Okubo,
Shinjuku-ku, Tokyo 169-8555, Japan}
\affiliation{$^2$ Advanced Research Institute for Science and Engineering,
Waseda  University, Shinjuku, Tokyo 169-8555, Japan}
\affiliation{$^3$ Waseda Institute for Astrophysics, Waseda University,
Shinjuku,  Tokyo 169-8555, Japan}

\date{\today}

\begin{abstract}
Studying Yang-Mills field and gravitational field in class A Bianchi spacetimes,
we find that chaotic behavior appears in the late phase
 (the  asymptotic future).  
In this phase, the Yang-Mills field behaves as that in Minkowski spacetime, 
in which we can understand it by a potential picture, except for the types VIII and IX.
At the same time, in the initial phase (near the initial singularity), 
we numerically find that the behavior seems to approach the Kasner solution.
However, we show that the Kasner circle is unstable and the Kasner solution is 
not an attractor.
From an analysis of stability and numerical simulation, we find a Mixmaster-like behavior in Bianchi I spacetime. 
Although this result may provide a counterexample to the BKL (Belinskii, Khalatnikov and Lifshitz)
conjecture, 
we show that the BKL conjecture is still valid in Bianchi IX spacetime.
We also analyze a multiplicative effect of two types of chaos, that is, 
chaos with the Yang-Mills field and that in vacuum Bianchi IX spacetime. 
Two types of chaos seem to coexist in the initial phase. 
However, the effect due to the Yang-Mills field is much smaller than 
that of the curvature term.
\end{abstract}

\maketitle

\section{Introduction}

The Yang-Mills field is one of the central objects in particle physics. 
When we discuss a unification of fundamental interactions, 
non-Abelian gauge fields,
which are a generalization of the Yang-Mills field, play a key role. 
Hence we should study more carefully the Yang-Mills field in cosmology, 
in particular in the early stage of the universe. 
If the field is random, it is described  by ``radiation", which is simple to 
deal with. 
The Yang-Mills field may also be homogeneously distributed 
by some symmetry-breaking mechanism.
So far, there has not been much study of the homogeneous Yang-Mills field 
in the context of cosmology.
In fact, the homogeneous source-free Yang-Mills field has been studied 
in the hope that a non-perturbative treatment may allow a better understanding 
of the vacuum state of the Yang-Mills field, despite the fact that strong and 
weak
interactions  have no classical counterpart. 
 It is also interesting to study the Yang-Mills field from the viewpoint 
of a dynamical system because it shows chaotic behaviors in Minkowski spacetime 
\cite{baseyna, matinyan(81), chirikov(81), chirikov(82), savvidy, spherical} 
(see also a good review \cite{Byro}). 
The homogeneous Yang-Mills field evolves as a particle in  some potential. 
A particle in such a potential 
(see Fig. \ref{graph:potential} for the potential shape)
is known to show chaotic behavior by the study of a billiard system.
The Yang-Mills field, however, is integrable in the Friedmann universe 
\cite{galtsov}.  
This result suggests that symmetry of the universe spoils its chaotic property
because a number of dynamical degrees of freedom decreases due to 
 the assumed symmetry \cite{C-couple}. 
In  axisymmetric Bianchi I spacetime, the Yang-Mills field again 
shows chaotic behavior \cite{darian,barrow}.  
Which spacetime shows the chaotic behavior of the Yang-Mills field?
Is chaos a universal phenomenon in generic spacetimes?
With these questions in mind, we search the Einstein-Yang-Mills  (EYM)
system 
for more generic Bianchi type spacetimes \cite{Belinskii}.

We should also stress another important point  in the present system.
We know that the past asymptotic behavior toward the initial
singularity of vacuum Bianchi VIII and IX spacetimes is chaotic 
\cite{BKL, Misner, Berger(94), Cornish, ML}. 
They are called the Mixmaster universe \cite{Misner}.  
The other type Bianchi spacetimes with a perfect fluid do not
show chaotic behavior. 
For example, a Bianchi I spacetime with a perfect fluid approaches 
monotonously the Kasner spacetime,  which is a shear dominant universe.  
The inclusion of a magnetic field, however, makes the dynamics very complicated. 
Bianchi I, II, VI$_0$ spacetimes with a perfect fluid and a homogeneous 
source-free magnetic field behave  like a Mixmaster universe 
\cite{leblanc(95), leblanc(97), leblanc(98)}. 
We should then study whether the past asymptotic behavior of Bianchi spacetimes  
with the Yang-Mills field shows similar chaotic behavior.

 We also mention the BKL (Belinskii, Khalatnikov and Lifshitz) conjecture, 
which predicts a fate of singularities in general inhomogeneous spacetimes 
\cite{BKL(82), Rendall}.
The BKL conjecture naively says that any spatially different points decouple 
each other
near initial singularity. 
Strictly speaking, it consists of the following three statements: \\
First, a spacetime near singularity becomes spatially homogeneous at each 
spacetime point, which can be described explicitly by  
 using a suitable local coordinate system. \\
Secondly, the contribution of a matter to the system becomes negligible 
near singularity compared with those of  shear and curvature terms, 
although the matter energy density  may blow up into initial singularity. 
This asymptotic spacetime is given by a vacuum Bianchi model.
This is why vacuum models are often studied.\\ 
Thirdly, the most generic spatially homogeneous models near singularity is 
Bianchi IX spacetime.
These statements, then, yields that a spacetime near initial singularity 
should be described by the Mixmaster universe. 
Now let us look back at the case of Bianchi I, II, VI$_0$ model 
with magnetic field \cite{leblanc(95), leblanc(97), leblanc(98)}.
In their cases, the energy density of magnetic field, 
which is normalized by a Hubble expansion parameter, 
does not vanish toward a singularity. 
It sometimes shows a spiky growth 
like a curvature term in the Mixmaster universe. 
This breaks the second statement of the BKL conjecture. 
Hence, such a model may be regarded as a counterexample to the BKL conjecture. 
It is interesting to study whether Bianchi I (or  more generically Bianchi IX) 
spacetime 
with the Yang-Mills field also give a counterexample to the BKL conjecture.

A natural question arises in the case of Bianchi IX spacetime:
Does a combination of two types of chaos, 
that is chaos with the Yang-Mills field and chaos in the Bianchi IX spacetime,
strengthen each chaotic behavior?
We now proceed to analyze the Yang-Mills field in the Bianchi IX spacetime.

Finally, it should be noted that it is difficult to discuss chaos in general 
relativity. 
Many standard indicators of chaos, such as the Lyapunov exponent, 
depend on the choice of coordinate systems. 
In order to define chaos independently from the coordinate choice, 
many efforts have been done so far. 
 A fractal basin approach is well 
known for the coordinate 
independent indicator of chaos \cite{Cornish, ML, barrow}. 
Using the Lyapunov exponent, recently, a coordinate independent definition of
 chaos has also been proposed \cite{Motter}. 
In this paper, we adopt a fractal basin approach to determine 
whether a system is chaotic or not.

We use the units of $c=\hbar=1$.

\section{basic equations}

The action of the EYM system is 
\begin{equation}
 S=\int
 d^4x\sqrt{-g}\left[\frac{1}{16\pi G}R
+\frac{1}{4g_{\rm YM}^2}F_{\mu\nu}^{(A)}F^{(A)\mu\nu}\right], 
\end{equation} 
where $R$ is the scalar curvature of the metric $g_{\mu\nu}$
and $F_{\mu\nu}^{(A)}$ is the field strength of Yang-Mills field.
``$A$'' describes the components of the  internal space.  The variation
of the action with respect to a metric 
$g_{\mu\nu}$ and a vector potential 
$A^{(A)}_{\mu}$ gives two basic equations, which
are the Einstein equations 
\begin{eqnarray}
 G_{\mu\nu}&=&8\pi G  T_{\mu\nu}, \\
 T_{\mu\nu}&=& \frac{1}{g_{\rm YM}^2}\left[F_{\mu\lambda}^{(A)}  F_{\
\nu}^{(A)\ \lambda}
 -\frac{1}{4}g_{\mu\nu}F_{\lambda\sigma}^{(A)} F^{(A)\lambda\sigma}\right]
\end{eqnarray}
and the Yang-Mills equations
\begin{equation}
 \vec{F}^{\mu\nu}_{\ \ \ ;\nu}-\vec{A}_{\nu}\times \vec{F}^{\mu\nu}=0\,,
\end{equation}
where $g_{\rm YM}$ is the self-coupling constant of the Yang-Mills field, 
and where $\vec{F}^{\mu\nu}=F^{(A) \mu\nu}\vec{\tau}_A$ and 
$\vec{A}_{\nu}=A_{\nu}^{(A)}\vec{\tau}_A$
with $\vec{\tau}_A$ being the SU(2) basis.
When we set our units as $8\pi G/g_{\rm YM}^2=1$ as well as $c=\hbar=1$,
the basic equations become free from the value of $g_{\rm YM}$.

We adopt the orthonormal-frame formalism, which has been developed in 
\cite{wainwright-ellis}.  In Bianchi-type spacetimes, there exists a
 3-dimensional homogeneous spacelike hypersurface
$\Sigma_t$ that is parameterized by a time coordinate  $t$. 
The timelike basis is given by $\bm{e}_0=\bm{\partial}_t$. 
The triad basis $\{\bm{e}_a\}$ on the hypersurface $\Sigma_t$ 
is defined by  the commutation function  $\gamma^c_{\ ab}$ as
\begin{equation}
[\bm{e}_a,\ \bm{e}_b]=\gamma^c_{\ ab} \bm{e}_c \,.
\end{equation}
It is convenient to decompose $\gamma^c_{\ ab}$ into the 
geometric variables (the Hubble
expansion
$H$, the acceleration
$\dot{u}_a$,  the shear $\sigma_{ab}$, and the  vorticity $\omega_{ab}$), 
and the variables
$\Omega_a$ (the rotation of $\bm{e}_a$), and the variables $a_a$
and $ n_{ab}$, which distinguish the type of Bianchi models, i.e., 
\begin{eqnarray}
 \gamma^a_{\ 0b} &=& -\sigma_{b}^{\ a} - H \delta_{b}^{\ a} 
                     - \epsilon^a_{\ bc} (\omega^c + \Omega^c) 
\label{gamma1} \\
 \gamma^0_{\ 0a} &=& \dot{u}_a \label{gamma2}\\
 \gamma^0_{\ ab} &=& -2\epsilon_{ab}^{\quad c} \omega_c \label{gamma3}\\
 \gamma^c_{\ ab} &=& \epsilon_{abd} n^{dc} + a_a \delta_b^{\ c} - a_b 
\delta_a^{\ c}\,. \label{gamma4}
\end{eqnarray}
In Bianchi spacetimes, the vorticity $\omega_{ab}$ always vanishes
because they are hypersurface-orthogonal. 

In order to analyze the EYM system, we have to write down the basic equations. 
For the generic situation, however, we have so many variables.
Hence we shall study the simplest case.
First, using the gauge freedom of the Yang-Mills field,
we set $A_0^{(A)}(t)=0$, which simplifies the vector potential as 
$\vec{\bm{A}}=A_a^{(A)}(t)\vec{\tau}_A \bm{\omega}^a$,  where
$\bm{\omega}^{\alpha}$ is  the dual basis of $\bm{e}_{\alpha}$ \cite{okuyama}. 

In the class A Bianchi type, $a_a$ vanishes. 
We can also  diagonalize $n_{ab}$, i.e. $n_{ab}=\mbox{diag} \{n_1, n_2, n_3\}$, 
using the remaining freedoms of a time-dependent rotation of the triad basis.
Then, we can show that if $\sigma_{ab},\ A_a^{(A)}$, and $\dot{A}_a^{(A)}$
do not have initially off-diagonal components, the equations of motion
guarantee that those variables always stay diagonal for class A
Bianchi spacetime (see Appendix). 
As a result, $\Omega_a$ vanishes, and 
the number of basic variables  reduces from 21 
(12 for spacetime [$H, N_a, \sigma_{ab}, \Omega_a$]  and 
9 for the Yang-Mills field [$A_a^{(A)}$]) 
to 10 
(7 for spacetime [$H, N_a, \sigma_{aa}$] and 3 for the Yang-Mills field 
[$Y_a^{(a)}$]).  
This is not valid for the class B Bianchi spacetimes. 
Therefore, in what follows, we discuss only the class A Bianchi spacetimes.

In order to discuss the dynamics, it is also convenient to introduce
the Hubble normalized variables, which are defined as follows:
\begin{eqnarray}
 \Sigma_{ab}\equiv 
\frac{\sigma_{ab}}{H},~~~\ N_{a}\equiv \frac{n_{a}}{H}\,.
\end{eqnarray}

We also introduce the conventional shear variables of the Bianchi
spacetime as
\begin{eqnarray}
\Sigma_+ \equiv \frac{1}{2}\left(\Sigma_{22}+\Sigma_{33}\right), ~~
\Sigma_-\equiv \frac{1}{2\sqrt{3}}\left(\Sigma_{22}-\Sigma_{33}\right),
~~\Sigma^2\equiv\Sigma_+^2+\Sigma_-^2 
\,.
\end{eqnarray}
The diagonal components of the Yang-Mills field potential are described by
new variables $a,b,c$ as
\begin{eqnarray}
 a \equiv A^{(1)}_1,\ b\equiv A^{(2)}_2,\ c\equiv A^{(3)}_3\,.
\end{eqnarray}
A new time variable $\tau$ defined by $d\tau=Hdt$ is introduced. 
$\tau$ denotes the e-folding number of the scale length. 
We choose the origin of $\tau$ when $H=1$.

Here we state the basic equations explicitly.
They consist of the generalized Friedmann equation, 
the dynamical Einstein equations, and the Yang-Mills equations. 
The generalized Friedmann equation, which is the Hamiltonian 
constraint equation, is described as
\begin{equation}
 \Sigma^2+\Omega_{\rm YM}+K=1 \,,\label{eq:FRW}  
\end{equation}
where 
$\Omega_{\rm YM}$ is the density parameter of the Yang-Mills field, i.e.  the
Hubble normalized energy density of the Yang-Mills field, which is given by
\begin{eqnarray}
 \Omega_{\rm YM} &\equiv& \frac{\mu_{\rm YM}}{3H^2} \nonumber\\
     &=&  \frac{1}{6} \left[
                      \left\{a'+(-2\Sigma_+ +1) a \right\}^2
                      +\left\{b'+(\Sigma_++\sqrt{3}\Sigma_-+1) b \right\}^2
                      +\left\{c'+ (\Sigma_+-\sqrt{3}\Sigma_-+1) c\right\}^2
\right.
\nonumber\\  
        & &  
\left.            +\left(N_1 a+{bc \over H}\right)^2
                      +\left(N_2 b+{ca \over H}\right)^2
                      +\left(N_3 c+{ab \over H}\right)^2
                       \right]\,,
\end{eqnarray}
 and where $K$ is the Hubble normalized curvature, which is defined by
\begin{eqnarray}
 K \equiv
-\frac{{}^{(3)}R}{6H^2}=
\frac{1}{12}\left\{N_1^2+N_2^2+N_3^2-2(N_1N_2+N_2N_3+N_3N_1)
\right\} \,.
\end{eqnarray}
Note that the positive 3-curvature corresponds to $K<0$. 
Then, except for the Bianchi type IX,
$K\geq 0$. 
The energy density of the Yang-Mills field is always positive definite. 
Thus we find that $\Sigma^2,\ \Omega_{\rm YM},$ and $K$ are restricted as
 $0\leq\Sigma^2\leq 1,\
0\leq\Omega_{\rm YM}\leq 1,$ and $\ 0\leq K\leq 1$, except for type IX. 

The dynamical Einstein equations  are
\begin{eqnarray}
H' &=& (-2-\Sigma^2+K)H \label{eq:H}\\ 
\Sigma_+' &=& (\Sigma^2-K-1)\Sigma_+-S_++\Pi_+ \label{eq:S+}\\
\Sigma_-' &=& (\Sigma^2-K-1)\Sigma_--S_-+\Pi_- \label{eq:S-}\\
N_1' &=& (\Sigma^2-K+1-4\Sigma_+)N_1 \label{eq:N1}\\
N_2' &=& (\Sigma^2-K+1+2\Sigma_++2\sqrt{3}\Sigma_-)N_2 \label{eq:N2}\\
N_3' &=& (\Sigma^2-K+1+2\Sigma_+-2\sqrt{3}\Sigma_-)N_3\,, \label{eq:N3}
\end{eqnarray}
where
\begin{eqnarray}
 S_+ &=& \frac{1}{6}\left\{(N_2-N_3)^2-N_1(2N_1-N_2-N_3)\right\}\\
S_- &=& \frac{1}{2\sqrt{3}}(N_3-N_2)(N_1-N_2-N_3)\,,
\end{eqnarray}
which depend on only $N_a$, and 
\begin{eqnarray}
 \Pi_+&=& -\frac{1}{6}
\left[
-2\left\{a'+(-2\Sigma_++1)a \right\}^2
+\left\{b'+(\Sigma_++\sqrt{3}\Sigma_-+1)b\right\}^2
+\left\{c'+(\Sigma_+-\sqrt{3}\Sigma_-+1)c\right\}^2
\right.
\nonumber\\ 
     & & 
\left.  
-2\left(N_1a+{bc\over H}\right)^2
+\left(N_2b+{ca\over H}\right)^2
+\left(N_3c+{ab\over H}\right)^2
\right]\\
\Pi_-&=& \frac{1}{2\sqrt{3}}
\left[
-\left\{b'+(\Sigma_++\sqrt{3}\Sigma_-+1)b\right\}^2
+\left\{c'+(\Sigma_+-\sqrt{3}\Sigma_-+1)c\right\}^2  
\right.
\nonumber\\ 
     & &  
\left.
-\left(N_2b+{ca\over H}\right)^2
+\left(N_3c+{ab\over H}\right)^2
\right]\,,
\end{eqnarray}
which are the Hubble normalized anisotropic pressures of the Yang-Mills field. 

The Yang-Mills field equations are
\begin{eqnarray}
 a''&=&(\Sigma^2-K-1)a' \nonumber\\
   & &+(\Sigma^2-5K+4\Sigma_+^2-4\Sigma_+-N_1(N_2+N_3)+2\Pi_+)a \nonumber\\
   & &-\frac{1}{H}(N_1+N_2+N_3)bc-\frac{1}{H^2}a(b^2+c^2)
\label{eq:a}\\
b''&=&(\Sigma^2-K-1)b' \nonumber \\
   & &+(\Sigma^2-5K+(\Sigma_++\sqrt{3}\Sigma_-)(\Sigma_++\sqrt{3}\Sigma_-+2)
   -N_2(N_3+N_1)-\Pi_+-\sqrt{3}\Pi_-)b \nonumber\\
   & &-\frac{1}{H}(N_1+N_2+N_3)ca-\frac{1}{H^2}b(c^2+a^2)
\label{eq:b}\\
c''&=& (\Sigma^2-K-1)c' \nonumber \\ 
   & & +(\Sigma^2-5K+(\Sigma_+-\sqrt{3}\Sigma_-)(\Sigma_+-\sqrt{3}\Sigma_-+2)
 -N_3(N_1+N_2)-\Pi_++\sqrt{3}\Pi_-)c \nonumber\\
   & & -\frac{1}{H}(N_1+N_2+N_3)ab-\frac{1}{H^2}c(a^2+b^2). 
\label{eq:c}
\end{eqnarray}

The basic variables are
\{$H,\ \Sigma_+,\ \Sigma_-,\ N_1,\ N_2,\ N_3,\ a,\ b,\ c$\}, whose equations  are
given by Eqs. (\ref{eq:H}), (\ref{eq:S+}), (\ref{eq:S-}), (\ref{eq:N1}),
(\ref{eq:N2}), (\ref{eq:N3}),  (\ref{eq:a}), (\ref{eq:b}), (\ref{eq:c}), with  one
constraint equation (\ref{eq:FRW}).

\begin{figure}[ht]
 \includegraphics[height=5cm]{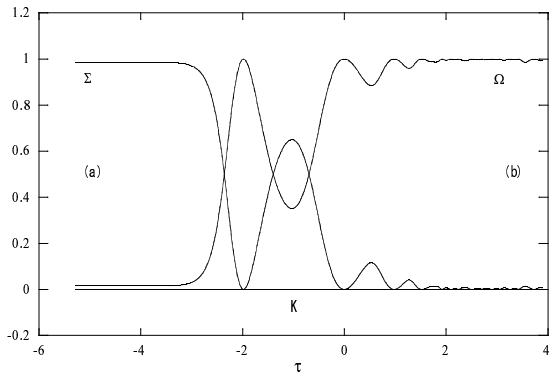} 
\caption{Two typical phases in the Bianchi I spacetime: 
(a) one is the initial phase ($\Sigma$-dominance) and 
(b) the other is the late phase   ($\Omega_{\rm YM}$-dominance), where
 $\Sigma$, $\Omega_{\rm YM}$, and $K$ denote the Hubble normalized shear strength, 
the Hubble normalized energy
 density of Yang-Mills field,
and the Hubble normalized curvature, respectively. These are constrained
by the generalized Friedmann equation: $\Omega_{\rm YM}+\Sigma^2+K=1\,.$} 
\label{graph:each_phase}
\end{figure}

For all class A Bianchi spacetimes except for type IX, 
we numerically find  two typical phases in the evolution of the universe;
i.e., an initial phase near a big-bang singularity, 
and a late phase, when a potential picture for the Yang-Mills field becomes valid. 
We will show typical behaviors in each phase
in Fig. \ref{graph:each_phase}, including a transient phase.
In what follows, we discuss the details of each phase in order.

\section{The late phase}

 In an axisymmetric Bianchi I model, Darian and K\"{u}nzle show that 
chaos appears in the dynamics  of the Yang-Mills field \cite{darian}. 
They adopt a potential picture to understand its chaotic behaviors.
Can we adopt the some approach to analyze the present system? 
To compare our analysis with their
results,  we shall transform our tetrad basis  to the Cartesian coordinate basis. 
The Yang-Mills field potential
$a, b, c$ should be replaced with
$A, B, C$ as
\begin{eqnarray}
 A\equiv \ell a\,,\ ~~
 B\equiv \ell b\,,\ ~~
 C\equiv \ell c\,,
\end{eqnarray}
where $\ell=\ell(t)$ is a scale factor defined by $H=\dot{\ell}/\ell$  and
given  explicitly as $\ell(t) = \ell_0 e^{\tau} = e^{\tau}$. 
$\ell_0$ is the initial value of the scale factor at $\tau=0$.   
Without a loss of generality, we can set $\ell_0=1$ by rescaling 
because the Hubble-normalized curvature $K$ does not depend on $\ell(t)$.

\begin{figure}[ht]
\begin{center}
\includegraphics[height=5cm]{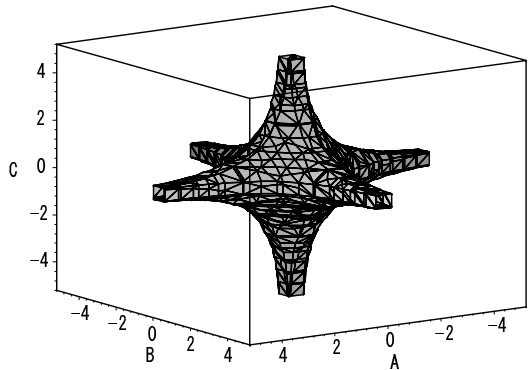}
\caption{The contour of the Yang-Mills field potential $V(A,B,C)$}
 \label{graph:potential}
\end{center}			   
\end{figure}

When we discuss the dynamics of the Yang-Mills field, adopting a potential picture
makes it convenient to see the existence of chaos, just as in Darian and K\"{u}nzle. 
For this purpose, we change our time coordinate from $\tau$
to a conformal time $\eta$, which is defined by 
$d\eta = \ell(t)^{-1}dt= H^{-1} e^{-\tau}d\tau$. 
Using this new coordinate, the Yang-Mills equations are written as 
\begin{eqnarray}
{ d^2A\over d\eta^2}&=& H^2 e^{2\tau}(-4K+4\Sigma_+^2-4\Sigma_+-N_1(N_2+N_3)
+2\Pi_+)A \nonumber\\ 
& &-H e^{\tau}(N_1+N_2+N_3)BC-{\partial V\over \partial A},\label{eq:A_eta} \\ 
{ d^2B\over d\eta^2}&=& H^2 e^{2\tau}(-4K+(\Sigma_++\sqrt{3}\Sigma_-)(\Sigma_++\sqrt{3}\Sigma_-+2)
   -N_2(N_3+N_1)-\Pi_+-\sqrt{3}\Pi_-)B \nonumber\\
& &-H e^{\tau}(N_1+N_2+N_3)CA -{\partial V\over \partial B}\label{eq:B_eta}\\ 
{ d^2C\over d\eta^2}&=& H^2 e^{2\tau}(-4K+(\Sigma_+-\sqrt{3}\Sigma_-)(\Sigma_+-\sqrt{3}\Sigma_-+2)
   -N_3(N_1+N_2)-\Pi_++\sqrt{3}\Pi_-)C \nonumber\\
& &-H e^{\tau}(N_1+N_2+N_3)AB -{\partial V\over \partial C}\label{eq:C_eta}\,,
\end{eqnarray}
where a potential $V$ is defined by
\begin{eqnarray}
 V(A,B,C)=\frac{1}{2}(A^2B^2+B^2C^2+C^2A^2) \label{eq:potential}\,,
\end{eqnarray}
whose shape is schematically shown in Fig. \ref{graph:potential}.
It is known that a test particle moving in such a potential 
shows a chaotic behavior
\cite{baseyna, matinyan(81), chirikov(81), chirikov(82), savvidy}.
Then, if we can adopt the potential picture, that is, 
if the potential term becomes dominant, 
we expect chaos in the dynamics of the Yang-Mills field. 

To show it, we  divide the r.h.s.  of Eqs. (\ref{eq:A_eta}), (\ref{eq:B_eta}), and
(\ref{eq:C_eta}) into three terms. 
The first one is proportional to the field itself ($A, B$, or $C$) 
and coupled to the  background  geometrical variables. 
The second one gives a coupling with other components of
the Yang-Mills field through $N_a$ (the variables   classifying Bianchi types).
The third one contains only the Yang-Mills  field variables. 
We then call these three terms an oscillation term, 
a Bianchi-mixing term, and  a potential term, respectively. 

If we can ignore the first and second terms, we can discuss the dynamics only 
via the potential $V$ and adopt a potential picture for the present dynamics.
In what follows, analyzing  the above three terms,
for each Bianchi model, we show that 
a potential term becomes dominant and a potential picture is valid 
in the late phase for some Bianchi models.

\subsection{Bianchi type I, II, and VI$_0$ spacetimes}

We first show that a potential picture is valid in the Bianchi I, II, and VI$_0$ 
spacetimes in the late phase by evaluating each term in Eqs.
(\ref{eq:A_eta}), (\ref{eq:B_eta}), and (\ref{eq:C_eta}).

\begin{itemize}
 \item Since $K\geq 0$,  $\Sigma^2\geq 0$, and $\Omega_{\rm YM}\geq 0$,
 from Eq. (\ref{eq:FRW}), we find that  $\Omega_{\rm YM}$, $\Sigma^2$, and $K$ are 
 bounded as $0\leq \Omega_{\rm YM} \leq 1$, $0\leq\Sigma^2 \leq 1$, and $0\leq K \leq 1$.
 $\Sigma_{\pm}$ is also restricted as $-1\leq \Sigma_{\pm} \leq 1$.

 \item Eq. (\ref{eq:H}) guarantees that $H$ damps exponentially. 
 Its exponent index is in the range of 
 $-1$ (curvature dominance) to $-3$ (shear dominance). 
 
 \item $\Pi_{\pm}$ is bounded as follows: \\
 Using $\Omega_{\rm YM}$, $\Pi_+$ is rewritten as
 \begin{eqnarray}
  \Pi_+ = -\Omega_{\rm YM} + \frac{1}{2}\left[ \{a'+(-2\Sigma_+ +1)a\}^2 
                                    + \left( N_1 a+\frac{1}{H}bc \right)^2 \right]\,,
 \end{eqnarray}
 which gives the constraint of $-1\leq\Pi_+$. $\Pi_+$ is also rewritten as
 \begin{eqnarray}
  \Pi_+ = 2\Omega_{\rm YM} - \frac{1}{2}\left[ \{b'+(\Sigma_+ +\sqrt{3} \Sigma_- +1)b\}^2 
                                    + \left( N_2 b +\frac{1}{H}ca \right)^2 \right. 
  \nonumber \\
                            \left.  +\{c'+(\Sigma_+ -\sqrt{3} \Sigma_- +1)c\}^2 
                                    + \left( N_3 c +\frac{1}{H}ab \right)^2 \right]\,,
 \end{eqnarray}
 which gives another constraint of $\Pi_+$ as $\Pi_+\leq 2$. 
 As a result, $\Pi_+$ is bounded as $-1\leq\Pi_+\leq2$. 
 $\Pi_-$ is similarly bounded as $-\sqrt{3}\leq\Pi_-\leq\sqrt{3}$. 

 \item In the case of Bianchi types I, II, and VI$_0$, 
 we can show that $N_a$ is bounded, that is, 
 \begin{itemize}
  \item In Bianchi I spacetime, $N_a=0$. 
 
  \item In Bianchi II spacetime,  $K=\frac{1}{12}N_1^2$. 
  Hence $N_a$ is bounded as $-2\sqrt{3}\leq N_1 \leq 2\sqrt{3}$.

  \item In Bianchi VI$_0$ spacetime, $K=\frac{1}{12}(N_1-N_2)^2$, 
  and $N_1$ and $N_2$ have different signs, i.e., $N_1 N_2 <0$. 
  Hence, $N_a$ is bounded as $-2\sqrt{3}\leq N_1, N_2 \leq 2\sqrt{3}$.
 \end{itemize}
\end{itemize}

From the above discussions, 
we find that both the oscillation and the Bianchi-mixing terms 
in Eq. (\ref{eq:A_eta}) damp 
and then that the potential term dominates with time,  
if the Yang-Mills field $A, B$ and $C$ do not decay exponentially.  
There is one exceptional case, in which $H$ decays exactly as $e^{-\tau}$.
Hence, except for such a special case, 
the Yang-Mills field behaves asymptotically ($\tau \rightarrow \infty$) 
as a particle moving in the potential $V(A,B,C)$. 
This potential shape is the same as that in Minkowski spacetime. 
Then, the Yang-Mills field will show a chaotic behavior just as 
in Minkowski spacetime \cite{savvidy}. 

\begin{figure}[ht]
\begin{tabular}{ccc}
 \includegraphics[height=6cm]{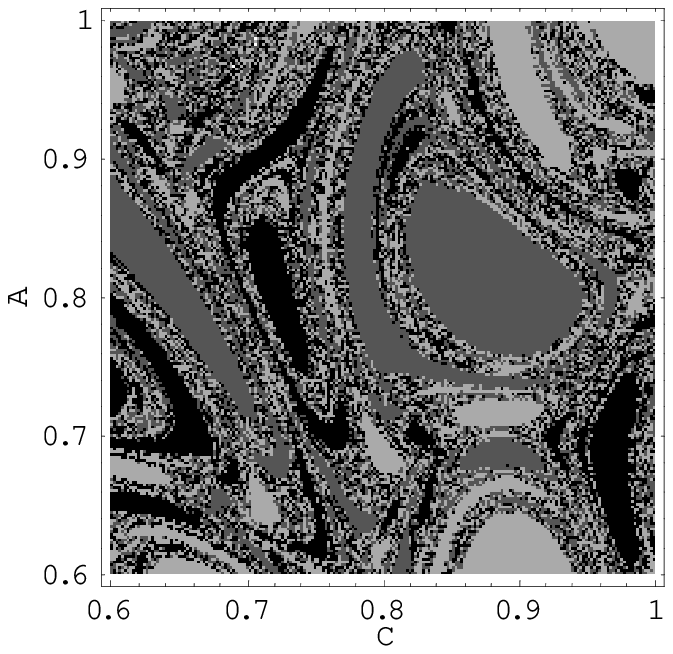} &
 \includegraphics[height=6cm]{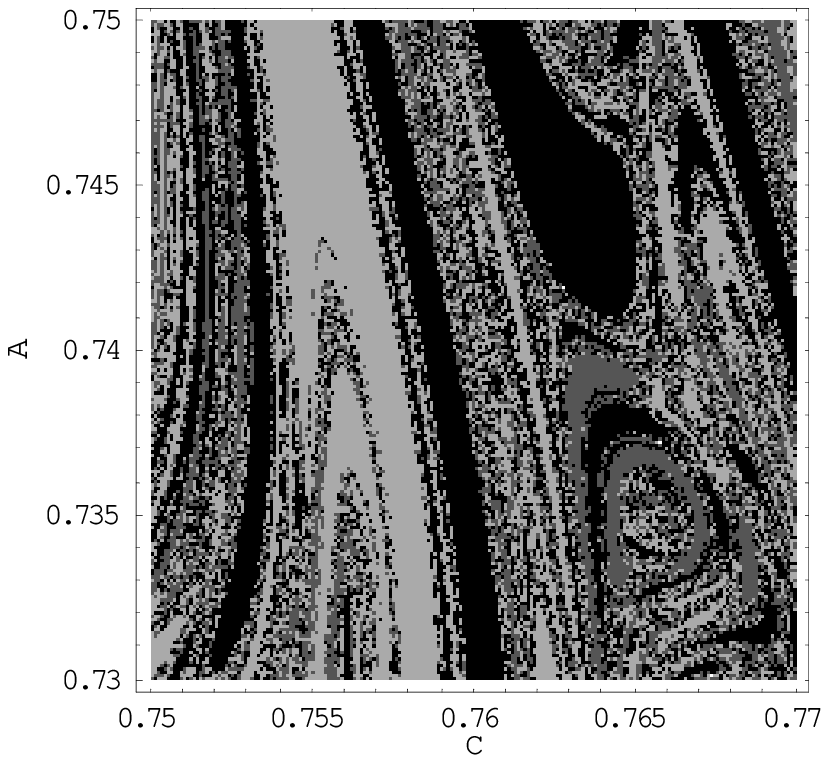} \\
 (a)&(b)\\
\end{tabular}
\caption{Fractal basin boundaries of initial data 
in Bianchi I spacetime. 
We set three windows at $|A|=2.8$ (bright gray),\ $|B|=2.8
$ (dark gray), and $|C|=2.8$ (black) in three channels
along $A$, $B$, and $C$ axes, where the fates of final state of orbits
are determined.  
Initial conditions are as follows: 
$\tau=0,\ H=1,\ \Sigma_+= 0.02,\ \Sigma_-=0.001,\ N_1=N_2=N_3=0,\ 
a'=0.3,\ b'=0.5,\ c'=0.7\,.$ Further, \\
(a) $A(=a)$ and $C(=c)$ vary from $0.6$ to $1.0\,.$\\
(b) An enlargement map of a part of Fig. {\ref{fig:fractal_YM}} (a) 
    [$0.73<A(=a)<0.75$ and $0.75<C(=c)<0.77$].\\
$B(=b)$ is obtained from Eq. (\ref{eq:FRW}). 
In each cases, a grid is $200\times 200$.} 
\label{fig:fractal_YM}
\end{figure}

In order to confirm the above results, we also analyze the present system numerically. 
To show its chaotic property, we show its sensitive dependence to initial data.
The potential (\ref{eq:potential}) has three channels along the $A,\ B$, and $C$ axes. 
We then set a window at a sufficient distance in each channel. 
When the orbit passes through one of these three windows, 
we assign three colors to the initial data in the following way: 
``bright gray," if the orbit passes through a window along $A$ axis; 
``dark gray," if it falls down in a window along the $B$ axis; 
and ``black" for a window along the $C$ axis. 
If numerical simulation is broken by increasing numerical errors 
before the orbit reaches any window, we assign ``white" to such initial data. 
We show our results in the Fig. \ref{fig:fractal_YM}.
Fig. \ref{fig:fractal_YM} shows that basin boundaries are fractal and 
orbits are sensitive to initial data, that is, orbits are chaotic. 
We find a relation between fractal basin boundaries and chaos similar 
to the other dynamical systems \cite{Cornish, ML}.

\begin{figure}[ht]
 \includegraphics[height=6cm]{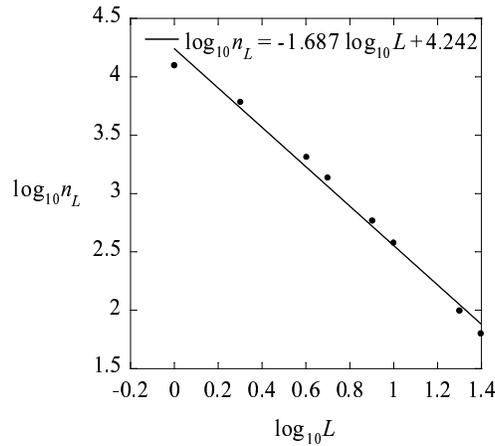}
\caption{ The fractal dimension of the ``black" region
in Fig. \ref{fig:fractal_YM} (a). 
The horizontal axis shows the logarithmic size of boxes ($\log_{10} L$), 
where $L$ is normalized by a grid size, 
while the vertical axis gives the logarithmic number of boxes ($\log_{10} n_L$) 
which contain black points. 
This figure shows the fractal dimension is $D=1.687$.
}
\label{fig:box}
\end{figure}

In order to see a strength of chaos, we 
calculate a fractal dimension of the basin boundaries of 
Fig. \ref{fig:fractal_YM}.
We adopt a box-counting method to evaluate the dimension.
We find that  the fractal dimension of the ``black" region
in Fig. \ref{fig:fractal_YM} (a) is $D=1.687$ 
(see Fig. \ref{fig:box}).

\subsection{Bianchi Type VII$_0$ and VIII spacetimes}

In Bianchi VII$_0$ and VIII spacetimes, it may be hard to show whether 
a potential picture is valid or not,  because $N_a$ is not bounded. 
We may, however, be able to  verify it by
evaluating the ``energy" $E$ numerically:
\begin{eqnarray}
 E=\frac{1}{2}A_{\eta}^2+\frac{1}{2}B_{\eta}^2+\frac{1}{2}C_{\eta}^2
  +\frac{1}{2}(A^2B^2+B^2C^2+C^2A^2).
\end{eqnarray}
If the potential picture is valid, $E$ will be asymptotically constant. 
We could successfully adopt the potential picture.
We shall then  evaluate $E$ in the case of Bianchi VII$_0$ and VIII 
spacetimes.

\begin{figure}[ht]
\begin{tabular}{ccc}
 \includegraphics[height=5cm]{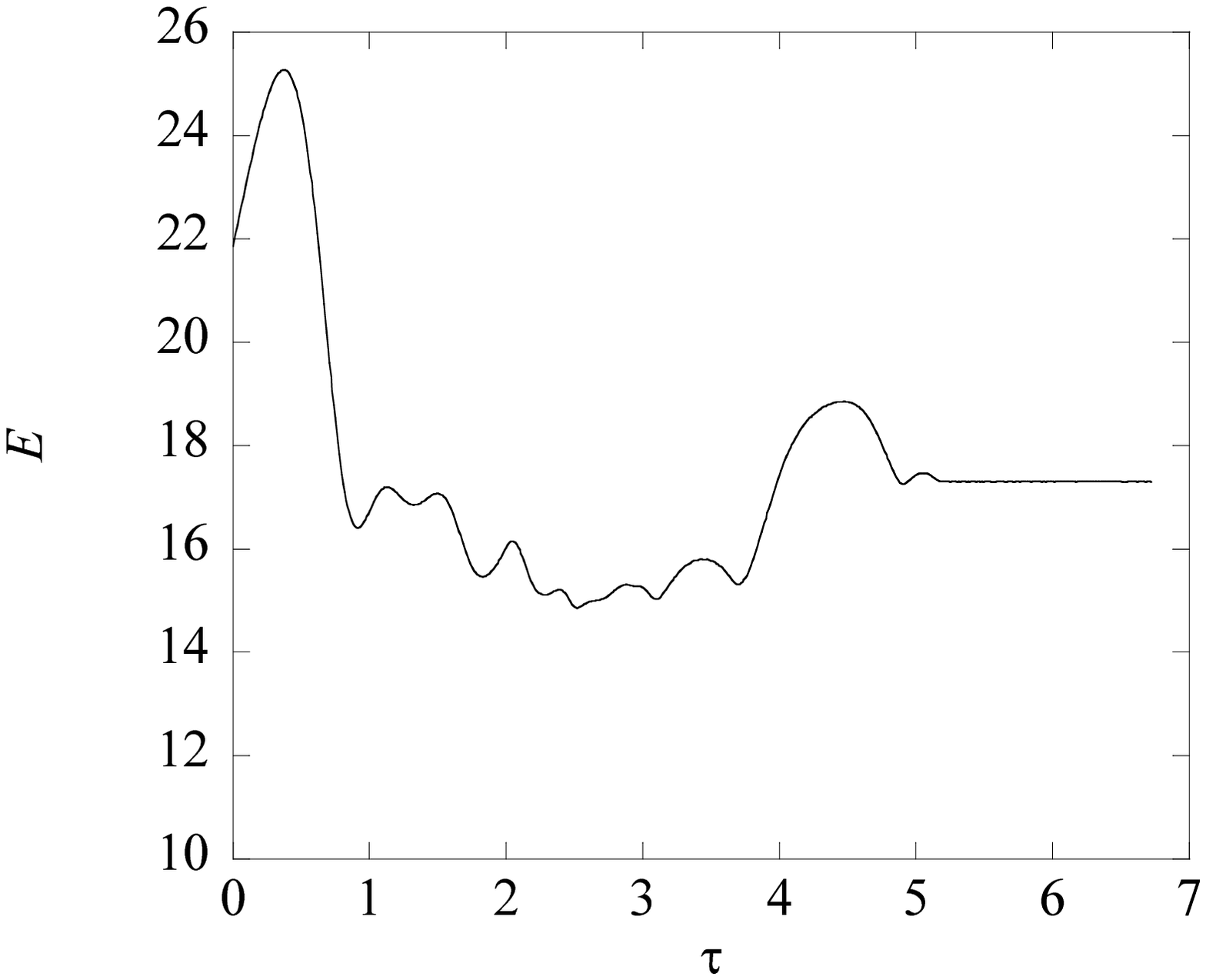} &
 \includegraphics[height=5cm]{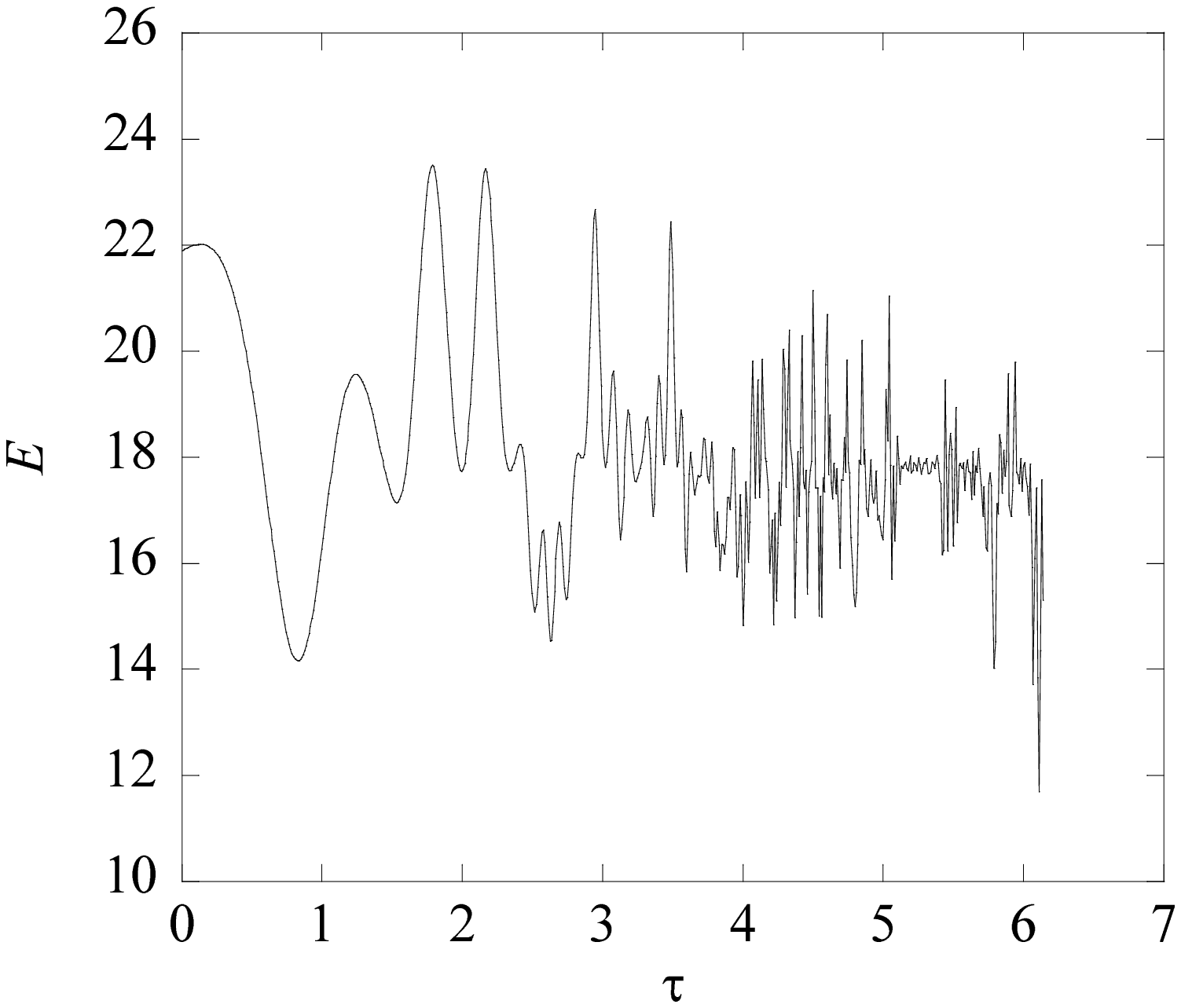} &
 \includegraphics[height=5cm]{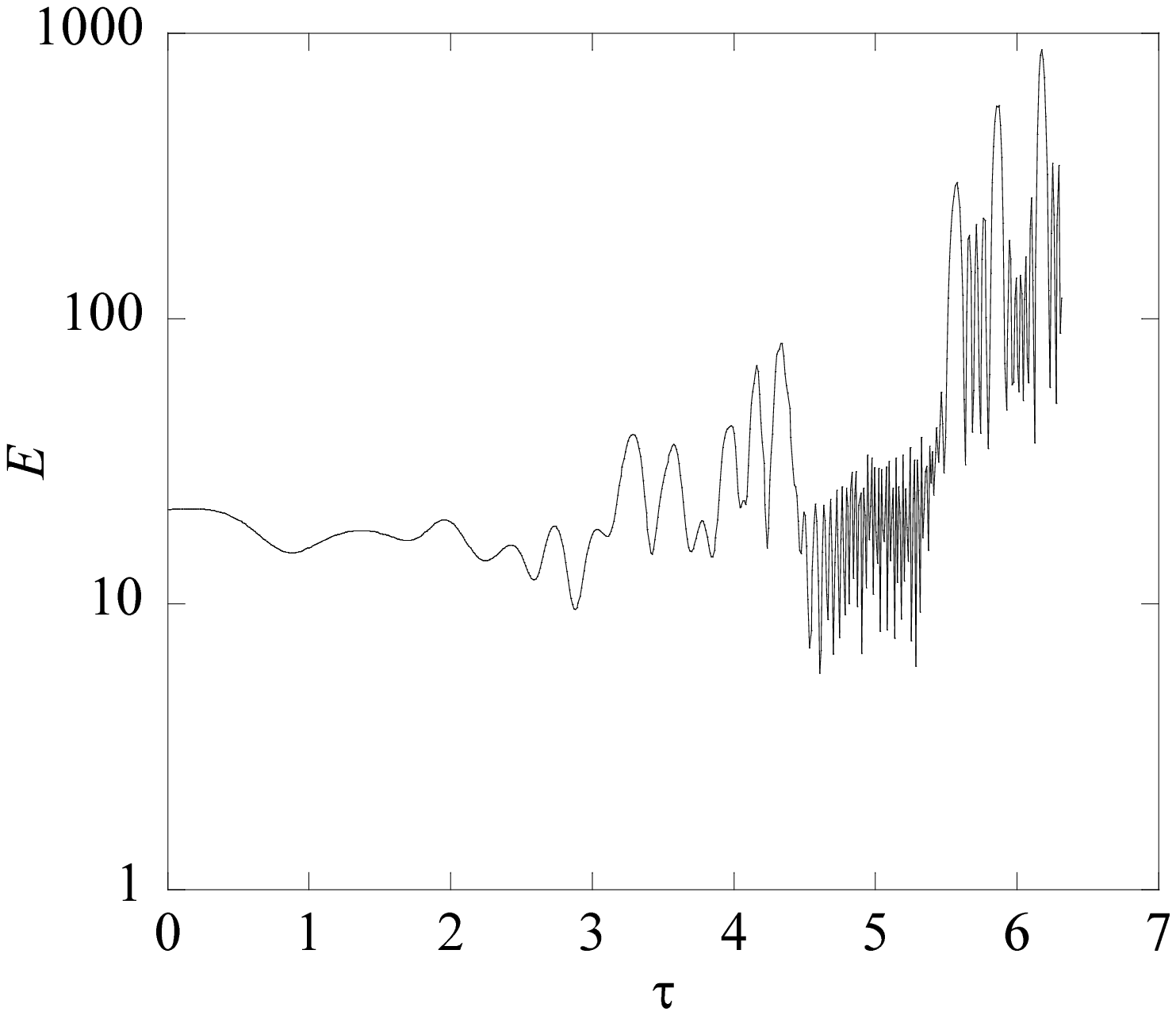} \\
 (a)&(b)&(c)\\
\end{tabular}
\caption{Time evolution of $E$: 
(a) Bianchi VI$_0$ spacetime. $E$ approaches constant for this chaotic system,
 in which a
potential picture is valid.  (b) Bianchi VII$_0$ spacetime.  
$E$ seems  to converge within
20\% fluctuations.  (c) Bianchi VIII spacetime. $E$ seems not to converge. } 
\label{fig:Hamiltonian}
\end{figure}

Fig. \ref{fig:Hamiltonian} (a) shows that $E$ approaches a constant 
for a Bianchi VI$_0$ model, 
which was shown to be chaotic in the previous subsection. 
In Bianchi VII$_0$ spacetime, $E$ seems to approach a constant; 
therefore a potential picture may be valid, 
even if  $N_a$ diverges (see Fig. \ref{fig:Hamiltonian} (b)). 
In Bianchi VIII spacetime, however,  if two of $N_a$ diverge,
then $E$ does not converge, and a potential picture may be no longer valid. 
However, we also find that when $N_a$ does not diverge, $E$ seems to approach 
constant, as in the case of Bianchi VII$_0$ spacetime.
It seems to depend on initial conditions whether $N_a$ diverges or not.
We could conclude that in Bianchi VIII, chaos via a Yang-Mills potential 
appears  for some initial conditions (i.e., $N_a$ is bounded), 
but a different type of dynamical behavior
is found for other initial conditions (i.e. $N_a$ diverges).
We are not sure whether the latter case is chaotic or not.

\begin{figure}[ht]
\begin{tabular}{ccc}
 \includegraphics[height=6cm]{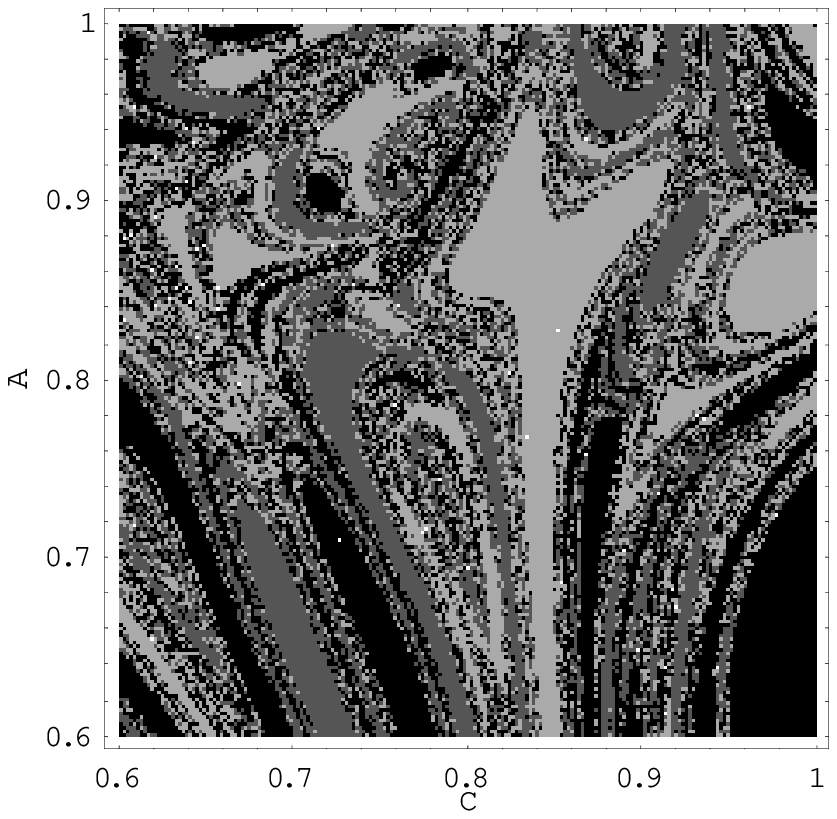} &
 \includegraphics[height=6cm]{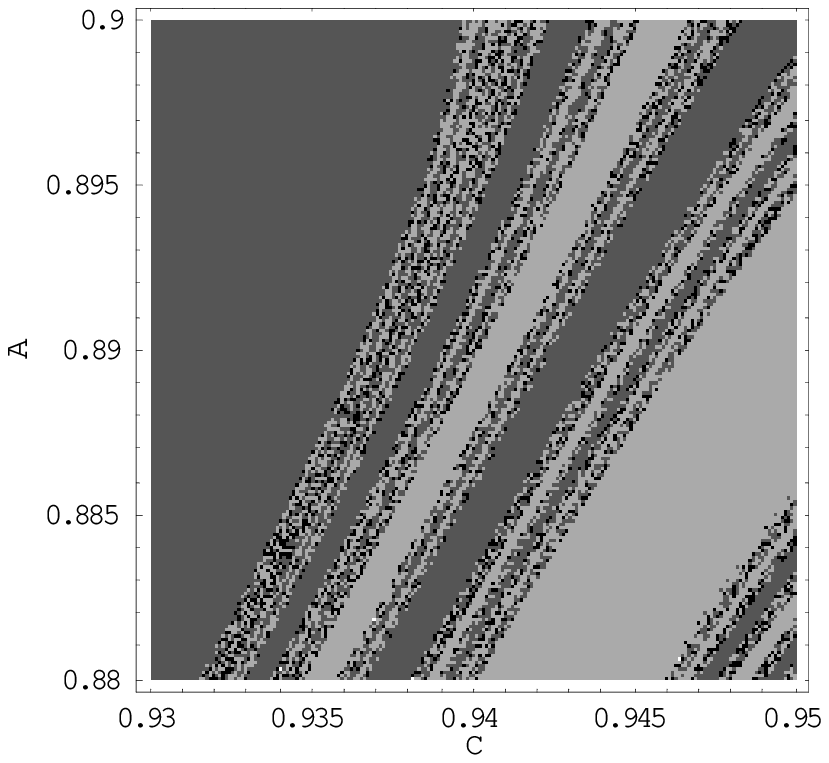} \\
 (a)&(b)\\
\end{tabular}
\caption{Fractal basin boundaries of initial data 
in Bianchi VII$_0$ spacetime. 
 We set three windows at $|A|=2.8$ (bright gray),\ $|B|=2.8
$ (dark gray), and $|C|=2.8$ (black) in three channels
along $A$, $B$, and $C$ axes, where the fates of final state of orbits
are determined.  Initial conditions are as follows: 
$\tau=0,\ H=1,\ \Sigma_+= 0.02,\ \Sigma_-=0.001,\ N_1=0.2,\ N_2=0.3,\ N_3=0,\ 
a'=0.3,\ b'=0.5,\ c'=0.7\,.$ Further, \\
(a) $A(=a)$ and $C(=c)$ vary from $0.6$ to $1.0\,.$\\
(b) An enlargement map of a part of Fig. (a) [$0.88<A(=a)<0.9$ and 
$0.93<C(=c)<0.95$].\\
$B(=b)$ is given from Eq. (\ref{eq:FRW}).
In each cases, a grid is $200\times 200$.} 
\label{fig:fractal_YM_70}
\end{figure}

We also analyze the sensitive dependence on initial data 
for Bianchi VII$_0$ and VIII spacetimes.
From  Fig. \ref{fig:fractal_YM_70}, we find that basin boundaries are fractal
and  
the box-counting dimension of the `` black" region 
in Fig. \ref{fig:fractal_YM_70} (a)
is $D=1.714$.
We conclude that the EYM system in the Bianchi VII$_0$ system is 
chaotic as expected from the evaluation of $E$. 

\begin{figure}[ht]
\begin{tabular}{cc}
 \includegraphics[height=6cm]{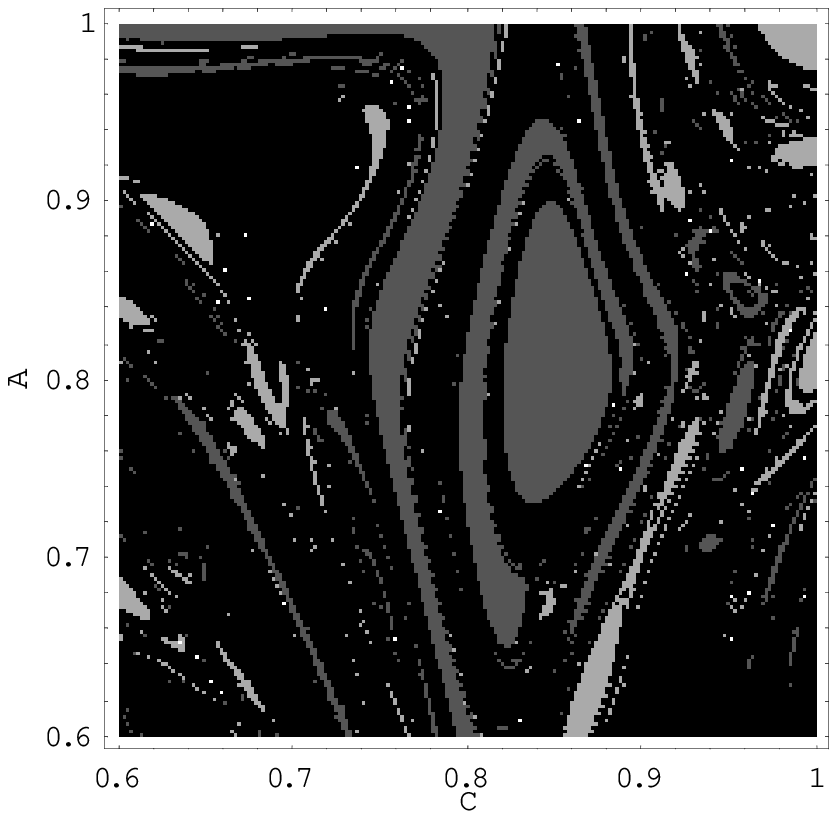} &
 \includegraphics[height=6cm]{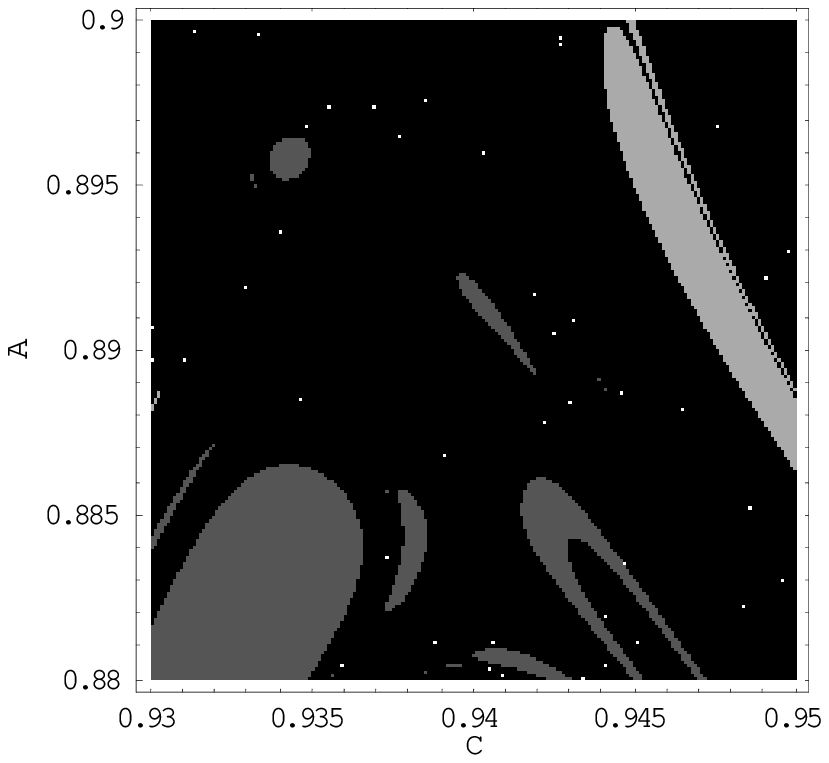} \\
 (a)&(b)\\
\end{tabular}
\caption{Fractal basin boundaries of initial data 
in Bianchi VIII spacetime. We set three windows at $|A|=2.8$ (bright gray),\ 
$|B|=2.8
$ (dark gray), and $|C|=2.8$ (black) in three channels
along $A$, $B$, and $C$ axes, where the fates of final state of orbits
are determined.  Initial conditions are the following: 
$\tau=0,\ H=1,\ \Sigma_+= 0.02,\ \Sigma_-=0.001,\ N_1=0.2,\ N_2=0.3,\ N_3=-0.2,\ 
a'=0.3,\ b'=0.5,\ c'=0.7\,.$ Further, \\
(a) $A(=a)$ and $C(=c)$ vary from $0.6$ to $1.0\,.$\\
(b) An enlargement map of a part of Fig. (a) [$0.88<A(=a)<0.9$ and 
$0.93<C(=c)<0.95$].\\
$B(=b)$ is given from Eq. (\ref{eq:FRW}). 
In each cases, a grid is $200\times 200$.} 
\label{fig:fractal_YM_8}
\end{figure}

On the other hand, Fig. \ref{fig:fractal_YM_8} (for Bianchi VIII) 
looks different 
from the previous fractal structures.  
Most parts of Fig. \ref{fig:fractal_YM_8} are black. 
The box-counting dimension of the ``black" region 
in Fig. \ref{fig:fractal_YM_8} (a) is $D=1.921$. 
The box-counting dimension is much closer to $D=2$ 
compared with the cases of  Bianchi I and VII$_0$.
Then  Fig. \ref{fig:fractal_YM_8} seems not to be fractal.
It seems  that the oscillation  and the Bianchi-mixing terms enlarge $C$, 
and then most orbits get into the $C$ window.  
It may depend on the values of $N_a$ which window is chosen by most orbits.  
Figs. \ref{fig:fractal_YM_70} and \ref{fig:fractal_YM_8} support our expectation which is obtained by asymptotic behavior of $E$ given 
in Fig. \ref{fig:Hamiltonian}. 

\subsection{Bianchi type IX spacetime}

In the case of Bianchi IX, the behavior is very similar to Bianchi VIII.
The most important difference is the asymptotic behavior of the spacetime.
Since Bianchi IX spacetime is closed,
once the  Hubble normalized variables diverge,
the Hubble parameter $H$ eventually vanishes
within a finite time, and the universe evolves into a big crunch. 
In that case, we cannot see anything about chaos in this period.

Let us take a more detailed look. 
If curvature term $K$ is negative, $H$ vanishes rapidly, so that 
Yang-Mills field $A,\ B,\ C$ do not have enough time to oscillate many times. 
We cannot determine whether such a system is chaotic or not. 
If one of $N_a$ is much smaller than the others, e.g., $N_3 \ll N_1,N_2$, however, 
$K$ becomes positive as $K\approx (N_1-N_2)^2/12$, 
and then $H$ may not vanish so rapidly, so that 
the Yang-Mills field $A,\ B,\ C$ could have enough time to oscillate many times. 
If one of $N_a$ is dominant, $N_a$ is bounded as $N_a\le 2\sqrt{3}$, 
using the generalized Friedmann equation (\ref{eq:FRW}) as long as $K$ is positive.
In this case, $E$ seems to approach a constant value, 
just as in the case of the Bianchi VIII spacetime, so that a potential picture holds. 
Although the spacetime eventually recollapses into a big crunch, 
we can find a temporal chaos for some period. 
However, depending on its initial conditions, we may
find divergence of $N_a$, and then a negative $K$ in the evolution of the spacetime.
In this case, we will find a big crunch before the appearance of
chaos in the Yang-Mills field.

When a potential picture is valid, the oscillation term and the Bianchi term 
turn out to be negligible. 
In this case, there is not much difference between Bianchi I and IX spacetimes. 
We conclude that there is no enhancement 
effect by spacetime chaos even in Bianchi IX spacetime
because the vector potential behaves as in Minkowski spacetime.

\section{the initial phase}

\subsection{Bianchi type I spacetime}
From Fig. \ref{graph:each_phase}, we find that the energy density of the Yang-Mills
field ($\Omega_{\rm YM}$) is damping to zero in the past direction near the initial
singularity.  
The  Bianchi I EYM system seems to approach the vacuum Bianchi I system, 
i.e., the Kasner solution.
The Kasner solution is described by $\Sigma^2=1$ setting  $\Omega_{\rm YM}=K=0$, 
which is called the Kasner circle. 
The Kasner circle is a past attractor in the case of Bianchi I spacetime 
with a non-tilted perfect fluid $0\leq\gamma\leq2$ \cite{wainwright-ellis}, 
where $\gamma$ is a parameter of an equation of state $p=(\gamma-1)\mu$. 
However, we have to be careful in drawing conclusions from our numerical analysis.
In the case of the Bianchi IX model, the evolution is described by a sequence of 
Bianchi I, i.e. the spacetime first approaches to some Bianchi I spacetime but it 
bounces via a Bianchi IX potential back to another Bianchi I spacetime. This 
process repeats infinite times to the initial singularity.  
Even in the Bianchi I model,  if we include a magnetic field, 
it is shown that the Kasner circle  has a two-dimensional unstable manifold 
in the past direction, just as in the Bianchi IX model \cite{leblanc(97)}. 
So it is important to study
whether the Kasner circle is a past attractor or not in our Bianchi I EYM system.

First, we consider the case of the Bianchi I model. 
From Fig. \ref{graph:each_phase}, 
one may  anticipate that the Kasner circle is a past attractor.  
Suppose that the spacetime approaches to one Kasner solution, that is 
$\ K= 0,\ \Omega_{\rm YM}\approx 0$, $\Psi \rightarrow \psi$ (some constant), and
$\Sigma\rightarrow 1$ (the Kasner circle), where 
$\Sigma_+=\Sigma\cos\Psi,\ \Sigma_-=\Sigma\sin\Psi$.
Near the Kasner circle, we obtain the following equation
\begin{eqnarray}
 \Omega_{\rm YM}&=&\frac{1}{6}\left[\{a'-(2\Sigma\cos \Psi-1)a\}^2
 +\left\{b'-\left(2\Sigma\cos\left(\Psi+\frac{2}{3}\pi\right)-1\right)b\right\}^2
  \right.                                                
+\left\{c'-\left(2\Sigma\cos\left(\Psi-\frac{2}{3}\pi\right)-1\right)c\right\}^2
 \nonumber \\
 & &
  \left. 
+\left(N_1 a+\frac{1}{H}bc\right)^2
+\left(N_2 b+\frac{1}{H}ca\right)^2
+\left(N_3 c+\frac{1}{H}ab\right)^2\right]\,, 
\label{eq:Omega}
\end{eqnarray}
which yields
\begin{eqnarray}
 a'=(2{\Sigma}\cos\Psi-1)a,\
 b'=\left\{2{\Sigma}\cos\left(\Psi+\frac{2}{3}\pi\right)-1\right\}b,\ 
 c'=\left\{2{\Sigma}\cos\left(\Psi-\frac{2}{3}\pi\right)-1\right\}c. 
\label{eq:exp_index}
\end{eqnarray}
These equations lead to  a situation in which at least one variable of {a,b,c} 
diverges exponentially in the limit of Bianchi I spacetime ($\Sigma=1$ and $\Psi=\psi$)
and one variable at least  decays exponentially in the past direction. 
Using Eq. (\ref{eq:exp_index}), we evaluate the last terms in Eq. (\ref{eq:Omega}),
\begin{eqnarray}
 \frac{bc}{H}& \propto&\exp\left[\{2\cos(\psi+\pi)+1\}\tau \right]\nonumber \\
 \frac{ca}{H} &
\propto&\exp\left[\left\{2\cos\left(\psi-\frac{1}{3}\pi\right)+1\right\}\tau\right] 
\nonumber \\
 \frac{ab}{H}& \propto&
\exp\left[\left\{2\cos\left(\psi+\frac{1}{3}\pi\right)+1\right\}\tau \right]\,.
\label{eq:YM_growth_rate}
\end{eqnarray}
If $-\pi<\psi<\frac{1}{3}\pi$,  
if $\frac{1}{3}\pi<\psi<\frac{5}{3}\pi$, and if
$-\frac{1}{3}\pi<\psi<\pi$, then ${ab}/{H}$,
${bc}/{H}$, and ${ca}/{H}$ damps to the past ($\tau \rightarrow -\infty$),
respectively. 
As a result, ${ab}/{H}$,
${bc}/{H}$, and ${ca}/{H}$ do not  damp simultaneously, 
and then $\Omega_{\rm YM}$ does not vanish in the past direction.  
This result is inconsistent with our initial assumption of $\Omega_{\rm YM}=0$.  
Hence we conclude that the Kasner circle is not a real past attractor, 
in other words, there is no asymptotic solution that approaches the Kasner circle. 

\begin{figure}[H]
\begin{tabular}{ccc}
 \includegraphics[height=5cm]{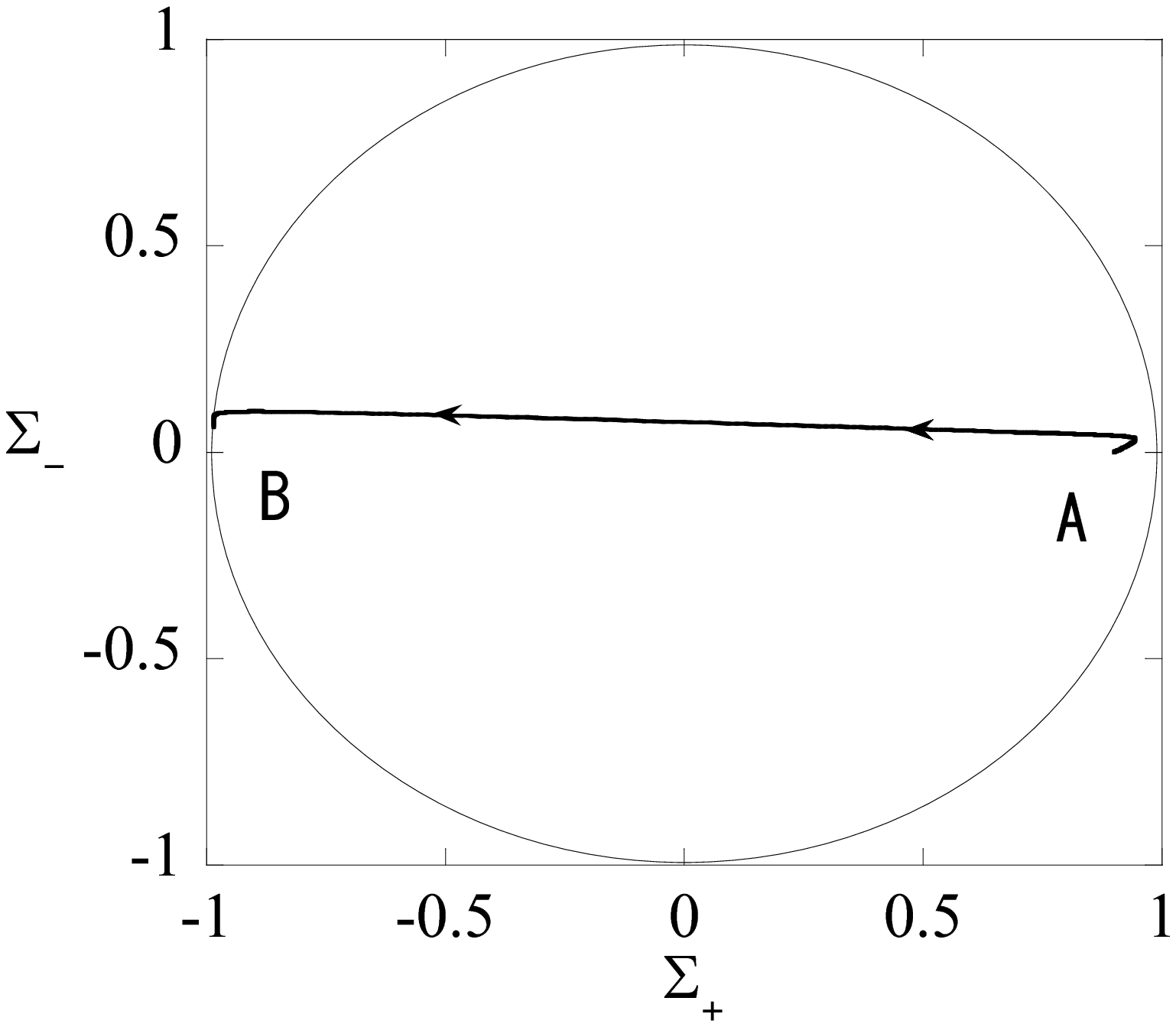} &
 \includegraphics[height=5cm]{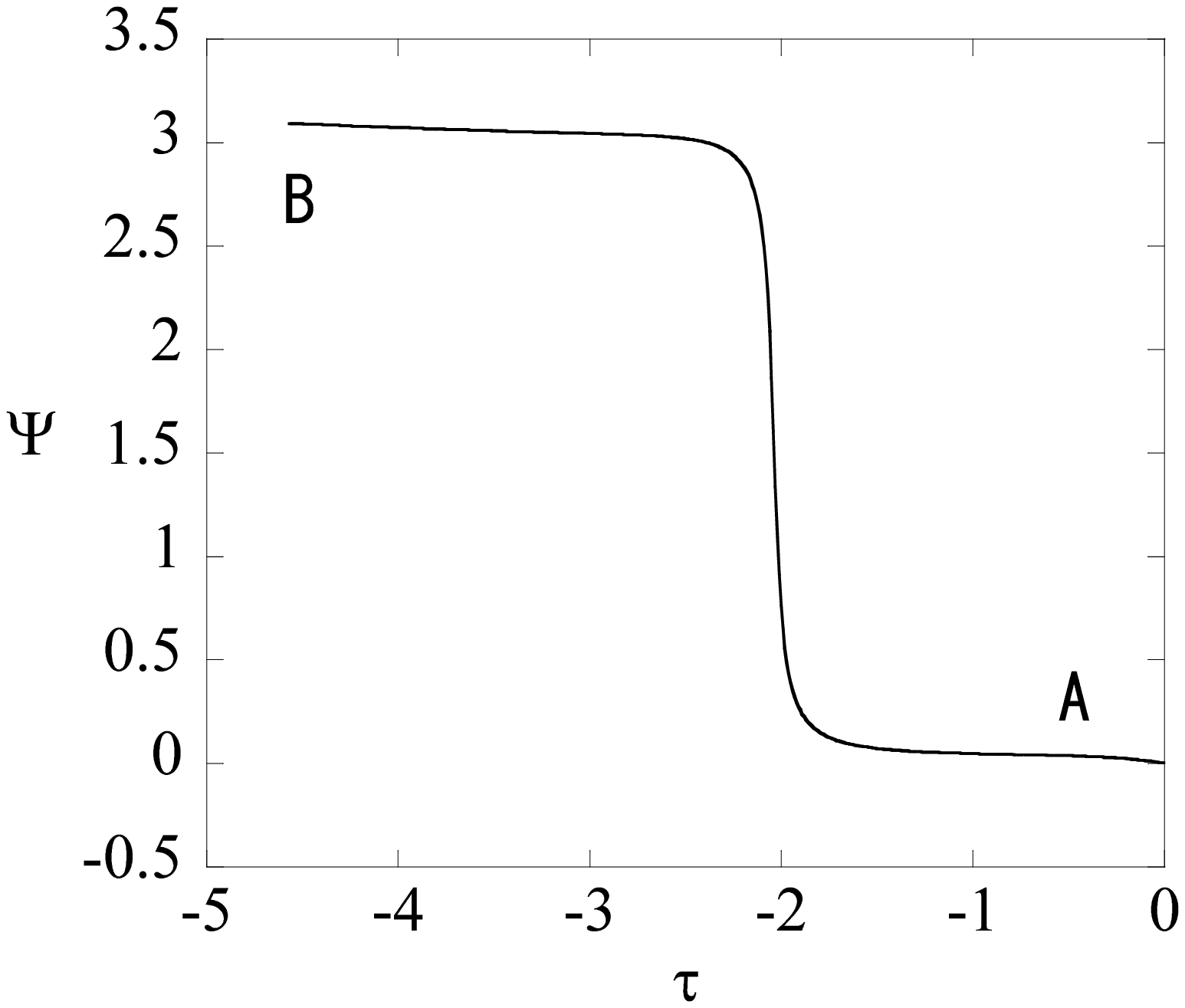} &
 \includegraphics[height=5cm]{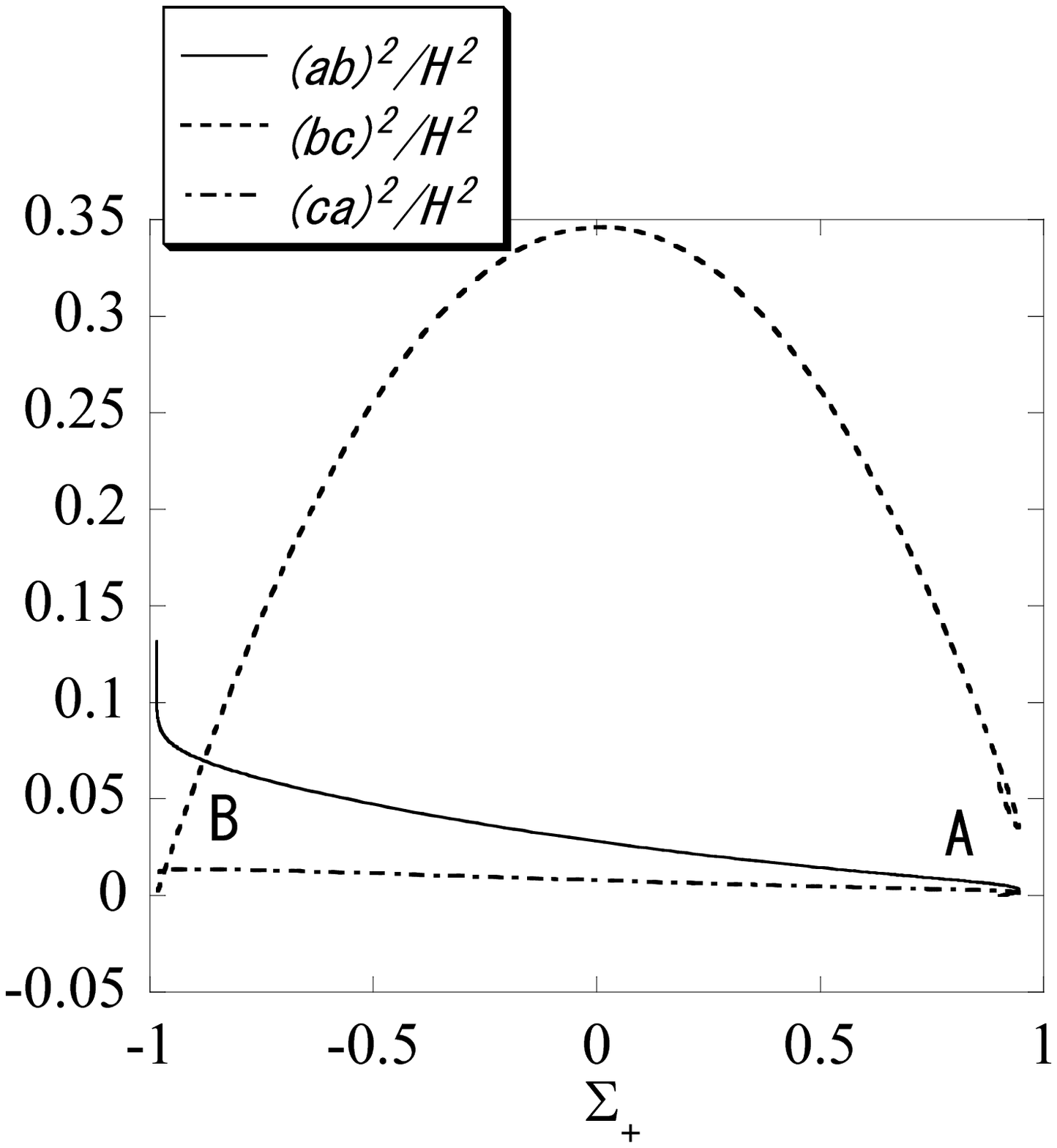} \\
 (a)&(b)&(c)\\
\end{tabular}
\caption{An orbit which starts near some point A on the Kasner circle 
evolves into another point B on the circle in the past direction: \\
 (a) The orbit in the $\Sigma_+$-$\Sigma_-$ plane. 
The initial values are $\Sigma=0.81$ and $\Psi=0$ (point A), 
which is close to the Kasner circle (a solid circle).
$\Psi$ eventually goes close to $\pi$  
(point B, which is  slightly less than $\pi$). \\
 (b) The time evolution $\Psi$ for the same orbit. 
The orbit endures for a long time near the Kasner circle,  
and eventually it moves to  another point on the  Kasner circle. \\
(c) The same orbit in the  $\Sigma_+$-$(ab)^2/H^2$, $\Sigma_+$-$(bc)^2/H^2$,
$\Sigma_+$-$(ca)^2/H^2$ planes.  
The initial values are $\Sigma=0.81$ and $(ab)^2/H^2=3.60\times10^{-7},
(bc)^2/H^2=5.76\times10^{-2}, (ca)^2/H^2=1.60\times10^{-7}$. 
$\Psi \sim 0$ results  in the initial growth of $(bc)^2/H^2$, 
but $\Psi \sim \pi$ (slightly smaller than $\pi$)  results in 
the decay of $(bc)^2/H^2$  and the growth of $(ab)^2/H^2$  in the final stage.}
\label{fig:Mixmaster_like}
\end{figure}

We have confirmed this result numerically. 
In Fig. \ref{fig:Mixmaster_like}, we depict the time evolution of $\Sigma_\pm$ 
and the  behavior of ${ab}/{H}$, ${bc}/{H}$, and ${ca}/{H}$.
In this figure, we start with the initial data near the Kasner solution 
(point A: $\Sigma=0.81$ and $\Psi=0$). 
Because Kasner spacetime has unstable modes, the spacetime does not approach 
the Kasner circle  ($\Sigma=1$), but rather $\Sigma$ is going to decrease 
(Fig. \ref{fig:Mixmaster_like} (b)) 
and $bc/H$ diverges (Fig. \ref{fig:Mixmaster_like} (c)).
Then the orbit again approaches another point of the Kasner circle.
We may expect that the same process will be repeated 
because the Kasner circle is not an attractor.

The mechanism preventing $\Omega_{\rm YM} \rightarrow 0$ is similar  to
that in  the Mixmaster solution.
In the case of the Mixmaster universe, when  a spacetime  approaching one point 
($\psi$) on the Kasner circle, one of $N_a$ grows while the others of $N_a$ damp.  
From the stability analysis of the Kasner circle in the vacuum Bianchi IX spacetime, 
the Kasner circle has a one-dimensional unstable manifold, that is,
 $N_1$ is unstable while $N_2, N_3 $ are stable 
when $-\frac{1}{3}\pi<\psi<\frac{1}{3}\pi$.
When a spacetime approaches other points on the Kasner circle as 
$\frac{1}{3}\pi<\psi<\pi$,
$N_2$ is unstable, while $N_3$ is unstable, if $-\pi<\psi<-\frac{1}{3}\pi\,.$  
These are described by the following equations:
\begin{eqnarray}
N_1 &\propto &
\exp \left[2\left\{2\cos\left(\psi+\pi\right)+1\right\}\tau\right] \nonumber \\
N_2  &\propto & 
\exp \left[2\left\{2\cos\left(\psi-\frac{1}{3}\pi\right)+1\right\}\tau\right]
\nonumber\\ 
N_3  &\propto &
\exp \left[2\left\{2\cos\left(\psi+\frac{1}{3}\pi\right)+1\right\}\tau\right]\,. 
\label{eq:BIX_growth_rate}
\end{eqnarray}

From Eqs. (\ref{eq:YM_growth_rate}) and (\ref{eq:BIX_growth_rate}),
we see that the Bianchi I EYM model is quite similar to the Mixmaster universe 
if we replace $N_1,\ N_2,$ and $N_3$ with ${bc}/{H}$, ${ca}/{H}$, and  
${ab}/{H}$. 
Therefore, although we cannot show the chaotic behavior in the EYM Bianchi I 
model in the
past direction, we conclude that such a spacetime does not approach the Kasner solution 
but that it probably shows a chaotic behavior, just as the Mixmaster 
universe does.
This indicates that energy density of the Yang-Mills field is not 
negligible 
compared to those of the shear and curvature terms. 
This breaks the second statement of BKL conjecture.

\subsection{Bianchi type IX spacetime}
Next, we consider the case of Bianchi IX spacetime. 
In this case we expect two types of chaos, i.e., the chaos of the Yang-Mills field 
and the chaos of spacetime. 
Do two types of chaos enhance each other, or evolve independently?
A study of the Bianchi IX EYM model by Belinskii and Khalatnikov \cite{Belinskii} 
gives hints on this question. 
They considered one Kasner epoch and found approximate solutions. 
Analyzing approximate solutions, it was shown that, 
although the spacetime is not affected by the Yang-Mills field between two Kasner epochs,
the Yang-Mills field changes sharply in the transition. 
These results anticipate that the chaos of spacetime does not depend on 
the Yang-Mills field. 
In our study, 
it is very difficult to judge because numerical error causes our simulation to stop 
just after a spacetime approaches some point on the Kasner circle.
Hence we shall take a guess by showing some indirect facts.
Eq. (\ref{eq:Omega}) yields Eq. (\ref{eq:YM_growth_rate}), 
when the spacetime approaches some point on the Kasner circle ($\Omega_{\rm YM}=0$). 
We also have 
\begin{eqnarray}
 (N_1 a)^2 &\propto& \exp[2\{2\cos(\psi+\pi)+1\}\tau] \nonumber \\
 (N_2 b)^2 &\propto& \exp\left[2\left\{2\cos\left(\psi-\frac{1}{3}\pi\right)+1
                                                               \right\}\tau\right] 
\nonumber \\
 (N_3 c)^2 &\propto& \exp\left[2\left\{2\cos\left(\psi+\frac{1}{3}\pi\right)+1
                                                               \right\}\tau\right] 
\label{eq:Omega_growth_rate}
\end{eqnarray}
and 
\begin{eqnarray}
 N_1^2 &\propto& \exp\left[4\left\{2\cos\left(\psi+\pi\right)+1\right\}\tau\right] 
\nonumber \\
 N_2^2 &\propto&
\exp\left[4\left\{2\cos\left(\psi-\frac{1}{3}\pi\right)+1\right\}\tau\right] \nonumber \\
 N_3^2 &\propto&
\exp\left[4\left\{2\cos\left(\psi+\frac{1}{3}\pi\right)+1\right\}\tau\right] \nonumber \\
 N_1 N_2 &\propto& \exp\left[4\left\{-\cos\left(\psi+\frac{1}{3}\pi\right)+1\right\}
\tau\right] \nonumber \\
 N_2 N_3 &\propto& \exp\left[4\left\{-\cos(\psi+\pi)+1\right\}\tau\right] \nonumber \\
 N_3 N_1 &\propto& \exp\left[4\left\{-\cos\left(\psi-\frac{1}{3}\pi\right)+1\right\}
\tau\right].
\label{eq:K_growth_rate}
\end{eqnarray}
These asymptotic behaviors show
that $\Omega_{\rm YM}$ diverges as 
$\max\left[(N_1 a)^2, (N_2 b)^2, (N_3 c)^2, (bc/H)^2, (ca/H)^2, (ab/H)^2 \right]$, 
while $K$ diverges as 
$\max\left[ N_1^2, N_2^2, N_3^2, N_1 N_2,  N_2 N_3,  N_3 N_1 \right]$. 
This means that even if a spacetime approaches one point on the Kasner circle once, 
it will leave the Kasner circle.
We expect that it may approach another point on the Kasner circle, 
just as in the vacuum Bianchi IX spacetime, 
mainly because of the divergence of $K$.
This is because the curvature term  $K$ diverges more strongly 
than the energy density of the Yang-Mills field $\Omega_{\rm YM}$.
The growth rate of $K$ is larger than that of $\Omega_{\rm YM}$.
The Yang-Mills field may not play any important role in this dynamics, 
which supports the results of Belinskii and Khalatnikov \cite{Belinskii}.
This means that the chaos of spacetime is stronger than 
that of the Yang-Mills field. 
This also means that the second statement of the BKL conjecture is satisfied,
because $\Omega_{YM}/K \rightarrow 0$ toward initial singularity. 
Considering that Bianchi IX spacetime is generic in spacelike homogeneous 
spacetime, 
we conclude that BKL conjecture is valid in the EYM system.

\section{conclusions} 

We studied the EYM system in class A Bianchi spacetime in order to generalize 
the result in axi-symmetric Bianchi I spacetime \cite{darian, barrow} 
and that of the Einstein-Maxwell system \cite{leblanc(95),leblanc(97),leblanc(98)}. 
We are also interested in the multiplicative effect of two types of chaos, 
that is, chaos found in Yang-Mills field and that found in Bianchi IX spacetime. 
We analyzed the case in which the shear and the Yang-Mills field can be diagonalized.
We find that  chaotic behavior  appears in the late phase (the  asymptotic future) 
for the Bianchi I, II, VI$_0$, and VII$_0$.  
In this phase, the Yang-Mills field behaves as that in Minkowski spacetime, 
in which we can understand it by a potential picture.
For types VIII and IX, a potential picture may no longer valid. 
Although we need further analyses, we find some sensitive dependence to initial data. 
It could be a new type of chaos in the Yang-Mills field.

While, in the initial phase (near the initial singularity), 
we numerically find that it seems to approach the Kasner solution.
However, we show that the Kasner circle is unstable and 
the Kasner solution is not an attractor.
From the analysis of stability and our numerical simulation, we expect that it behaves 
as a Mixmaster universe even for Bianchi I spacetime. 
 This result may provide a counterexample to the so-called BKL conjecture.
However, it is not the case for the Bianchi IX spacetime,
in which the BKL conjecture is still valid.

We also analyze a multiplicative effect of two types of chaos. 
Two types of chaos seem to coexist in the initial phase. 
However, the effect due to the Yang-Mills field is much smaller than 
that by the curvature term.

Finally we mention the transient phase. 
It is found that the behaviors of shear scalar $\Sigma^2$ 
and energy density of the Yang-Mills field $\Omega_{\rm YM}$ are very complicated. 
The potential picture of the Yang-Mills field is of course no longer valid. 
Hence this complexity is not only due to the chaos the Yang-Mills field 
appearing in the late phase. 
The chaos in the initial phase is understood by a mechanism 
similar to the Mixmaster universe. 
In the transient phase, however, the orbit does not need to approach the Kasner circle. 
So the behavior in the transient phase is not like the Mixmaster type. 
This could be new type of chaos that appears by a multiplicative effect, 
although it is a transient phenomena. 
This is under investigation.

\acknowledgments
We would like to thank J. Barrow for useful comments.
This work was partially supported by the Grant-in-Aid for Scientific Research
Fund of the MEXT (Nos. 14540281, 16540250 and 02041), by the Waseda
University Grant for Special Research Projects and a Grant for The 21st Century
COE Program (Holistic Research and Education Center for Physics
Self-organization Systems) at Waseda University.

\appendix

\section{Dynamics of off-diagonal components} 
\label{ch:appendix}

The vector potential of Yang-Mills field in a Bianchi spacetime is written as
$\vec{\bm{A}}=A_a^{(A)}(t)\vec{\tau}_A \bm{\omega}^a$, 
using the gauge freedom.
In this appendix, we show that if $\sigma_{ab},\ A_a^{(A)}$ and 
$\dot{A}_a^{(A)}$ are initially diagonal, 
the off-diagonal components will not appear anytime in class A Bianchi spacetimes, 
while they will appear in class B Bianchi spacetimes.

In class A Bianchi spacetime, each component of the Yang-Mills equation
$\vec{F}^{\alpha\beta}_{\ \ \ ;\beta} 
- \vec{A}_{\beta} \times \vec{F}^{\alpha\beta} = 0$
is written as follows:\\
$\alpha=0$  component :
\begin{eqnarray}
& & \vec{A}_1 \times \dot{\vec{A}}_1 +\vec{A}_2 \times \dot{\vec{A}}_2 
                                  +\vec{A}_3 \times \dot{\vec{A}}_3 
 -2\Omega_3 \vec{A}_1 \times \vec{A}_2 
    -2\Omega_1 \vec{A}_2 \times \vec{A}_3
    -2\Omega_2 \vec{A}_3 \times \vec{A}_1 =0;
\label{eq:a=0}
\end{eqnarray}
 $\alpha=1$ component :
\begin{eqnarray}
 \ddot{\vec{A}}_1 &=& -3H\dot{\vec{A}}_1 + 2\Omega_3 \dot{\vec{A}}_2
                                         - 2\Omega_2 \dot{\vec{A}}_3
 \nonumber \\ 
& & +\{ -(\dot{\sigma}_{11}+\dot{H})+(\sigma_{11}-2H)(\sigma_{11}+H)
       +(\sigma_{12}+\Omega_3)^2 +(\sigma_{13}-\Omega_2)^2 -(n_1)^2 \}\vec{A}_1
 \nonumber \\ 
& & +\{  (\dot{\Omega}_3-\dot{\sigma}_{12})
       -(\sigma_{11}-2H)(\sigma_{12}-\Omega_3)
       +(\sigma_{12}+\Omega_3)(\sigma_{22}+H)
       +(\sigma_{13}-\Omega_2)(\sigma_{23}+\Omega_1) \}\vec{A}_2 
 \nonumber \\ 
& & +\{ -(\dot{\Omega}_2+\dot{\sigma}_{13})
       +(\sigma_{11}-2H)(\sigma_{13}+\Omega_2)
       +(\sigma_{12}+\Omega_3)(\sigma_{23}-\Omega_1) 
       +(\sigma_{13}-\Omega_2)(\sigma_{33}+H) \}\vec{A}_3
 \nonumber \\
& & -(n_1+n_2+n_3)\vec{A}_2 \times \vec{A}_3
    -\vec{A}_2 \times (\vec{A}_1 \times \vec{A}_2)
    -\vec{A}_3 \times (\vec{A}_1 \times \vec{A}_3); 
\label{eq:Y-M_1}
\end{eqnarray}
and $\alpha=2 [{\rm or} ~3]$ component is obtained by exchanging indices in
Eq. (\ref{eq:Y-M_1}) 
as $(1,2,3)\rightarrow (2,3,1)[{\rm or} ~(3,1,2)]$. In class A Bianchi spacetime,
some Jacobi identities for $\Vec{e}_a$  give a relation between $\Omega_a$ and
$\sigma_{ab}$ as
\begin{eqnarray}
 \Omega_1=\frac{n_2+n_3}{n_2-n_3} \sigma_{23} \nonumber \\
 \Omega_2=\frac{n_3+n_1}{n_3-n_1} \sigma_{31} \nonumber \\
 \Omega_3=\frac{n_1+n_2}{n_1-n_2} \sigma_{12}\,.
\label{Omega-sigma}
\end{eqnarray}
The equation for the shear $\sigma_{ab}$ is given by
\begin{eqnarray}
 \dot{\sigma}_{ab}=-3H\sigma_{ab}+2\epsilon^{cd}_{\ \ (a} \sigma_{b)c} \Omega_d
                   -b_{ab}+\frac{1}{3}b_c^{\ c}\delta_{ab}+\pi_{ab},
\end{eqnarray}
where
\begin{eqnarray}
  b_{ab} \equiv 2n_a^{\ c}n_{cb}-n_c^cn_{ab}\,.  
\end{eqnarray}

We then assume $\sigma_{ab},\ A_a^{(A)}$ and $\dot{A}_a^{(A)}$ do not
have initially off-diagonal components. 
From Eq. (\ref{Omega-sigma}), $\Omega_a=0$. 
With this ansatz, Eq. (\ref{eq:a=0}) is trivial and Eq. (\ref{eq:Y-M_1}) is 
rewritten as : 
\begin{eqnarray}
 \ddot{\vec{A}}_1 &=& -3H\dot{\vec{A}}_1 
                      + \{ -(\dot{\sigma}_{11}+\dot{H})
                           +(\sigma_{11}-2H)(\sigma_{11}+H) 
                           -(n_1)^2 \}\vec{A}_1 \nonumber \\
                  & & -(n_1+n_2+n_3)A_2^{(2)} A_3^{(3)} \vec{\tau}_1
                      -A_1^{(1)} A_2^{(2)} A_2^{(2)} \vec{\tau}_1
                      -A_1^{(1)} A_3^{(3)} A_3^{(3)} \vec{\tau}_1, 
\label{eq:Y-M_2}
\end{eqnarray}
Note that  $\Omega_a=0$ from Eq. (\ref{Omega-sigma}). 
The equations for ${\vec{A}}_2 $ and ${\vec{A}}_3 $ are obtained by exchanging
indices.

Setting $\sigma_{ab}=0$ for $a\neq b$ and  $ A_a^{(A)}=0,
\dot{A}_a^{(A)}=0$ for
$a\neq A$, we obtain 
\begin{equation}
 \ddot{A}_{a}^{(A)}=0 \ \ ( a\neq A)\,.
\label{eq:YM_off}
\end{equation}
The off-diagonal components of dynamical equations for $\sigma_{ab}$ is reduced to 
\begin{equation}
 \dot{\sigma}_{ab}=\pi_{ab} \ \ (a\neq b)
\end{equation}
because off-diagonal components of $b_{ab}$ vanish, where $\pi_{ab}$ is the traceless
part of the energy momentum tensor of the Yang-Mills field, i.e.,
$\pi_{ab}\equiv T_{ab}-{1\over 3}T^c_c \eta_{ab}$ with
$T_{\alpha\beta}=\vec{F}_{\alpha\gamma}\vec{F}_{\beta}^{\gamma}
-\frac{1}{4}g_{\alpha\beta}\vec{F}^2$.
Writing down the off-diagonal part of  $\pi_{ab}$ explicitly, we can easily show  
$\pi_{ab}=0$ as follows:\\
\begin{eqnarray}
 \pi_{12} &=& \vec{F}_{1\gamma}\vec{F}_2^{\ \gamma} 
         = \vec{F}_{10}\vec{F}_2^{\ 0}+\vec{F}_{13}\vec{F}_2^{\ 3}
          \nonumber \\
          &=& -[\dot{\vec{A}}_1+(\sigma_{11}+H) \vec{A}_1]\cdot
               [\dot{\vec{A}}_2+(\sigma_{22}+H) \vec{A}_2] 
           +[n_2 \vec{A}_2 + A_1^{(1)}A_3^{(3)}\vec{\tau}_2]\cdot
               [-n_1 \vec{A}_1 - A_2^{(2)}A_3^{(3)}\vec{\tau}_1]
          \nonumber \\
          &=& 0.
\end{eqnarray}
We also find $\pi_{23}=0$ and $\pi_{31}=0$. Thus we obtain
\begin{equation}
 \dot{\sigma}_{ab}=0 \ \ (a \neq b )\,.
\label{eq:sigma_off}
\end{equation}

Eqs. (\ref{eq:YM_off})
and (\ref{eq:sigma_off})
guarantee that the off-diagonal components of
$\sigma_{ab},\ A_a^{(A)}$ and
$\dot{A}_a^{(A)}$ always vanish. 
We conclude that there is some set of initial
data   in class A Bianchi spacetime   for which $\sigma_{ab},\ A_a^{(A)}$
and
$\dot{A}_a^{(A)}$ are always diagonal.

Next we discuss the dynamics of off-diagonal components 
in class B Bianchi spacetimes.
For simplicity, we consider only the Bianchi V spacetime, in which $n_a=0$ and 
$a_a=(a,0,0)$.
The basic equations for Yang-Mills field are given as follows:\\
$\alpha=0$ component
\begin{eqnarray}
 & &2a(\dot{\vec{A}}_1-\Omega_3\vec{A}_2+\Omega_2\vec{A}_3
    +(\sigma_{11}+H)\vec{A}_1+\sigma_{12}\vec{A}_2+\sigma_{13}\vec{A}_3)  
-\vec{A}_1 \times \dot{\vec{A}}_1 -\vec{A}_2 \times \dot{\vec{A}}_2 
                                  -\vec{A}_3 \times \dot{\vec{A}}_3  
\nonumber \\
 & &+2\Omega_3 \vec{A}_1 \times \vec{A}_2 
    +2\Omega_1 \vec{A}_2 \times \vec{A}_3
    +2\Omega_2 \vec{A}_3 \times \vec{A}_1 =0; \label{eq:Y-M_0(V)}
\end{eqnarray}
and $\alpha=1$ component is given by Eq. (\ref{eq:Y-M_1}) with $n_a=0$.
 
$\alpha=2 [ {\rm or}~ 3]$ component is derived from the corresponding equation
in class A Bianchi spacetimes by inserting
$n_a=0$ and by adding the term 
\begin{eqnarray}
 a^2 \vec{A}_{2 [{\rm or}~3]}+2a \vec{A}_1 \times \vec{A}_{2 [{\rm or}~3]}.
\label{eq:additional_term}
\end{eqnarray}
Some Jacobi identities yield 
\begin{eqnarray}
 \dot{a}&=&-(H+\sigma_{11})a \\
 \Omega_2&=& \sigma_{31} \\
 \Omega_3&=& -\sigma_{12}.
\end{eqnarray}
The  equations for shear $\sigma_{ab}$ is
\begin{eqnarray}
 \dot{\sigma}_{ab}=-3H\sigma_{ab}+2\epsilon^{cd}_{\ \ (a} \sigma_{b)c} \Omega_d
                   +\pi_{ab} \,,
\label{eq:shear(V)}
\end{eqnarray}
where we assume $\sigma_{ab},\ A_a^{(A)}$ and $\dot{A}_a^{(A)}$ are diagonal. 
From Jacobi identities, we find $\Omega_2=\Omega_3=0$, but we do not know about $\Omega_1$. 
We then rewrite Eq. (\ref{eq:Y-M_0(V)}) as
\begin{eqnarray}
\Omega_1 (\vec{A}_2\times\vec{A}_3)
+a [\dot{\vec{A}}_1+(\sigma_{11}+H)\vec{A}_1] = 0\,.
\end{eqnarray}
Because the second term of the r.h.s. in this equation does not vanish
(unless $A_1^{(1)}=0$), we find
$\Omega_1\neq 0$, which results in   $\pi_{23} \neq 0$. 
Hence the source term of the off-diagonal components of Eq. (\ref{eq:shear(V)}) 
does not vanish and then $\sigma_{ab}$ becomes no longer diagonal.
The Yang-Mills equations show that $A_2^{(3)},\ \dot{A}_2^{(3)},\ 
A_3^{(2)}\,,$  and $\dot{A}_3^{(2)}$ become also non-trivial
due to the term (\ref{eq:additional_term}).  
We conclude that in type V Bianchi spacetime we cannot assume that 
$\sigma_{ab}, A_a^{(A)} $ and $\dot{A}_a^{(A)}$ are diagonal 
even though they do not have off-diagonal components initially.

\end{document}